\documentclass[manuscript]{acmart}

\AtBeginDocument{%
  }

\usepackage{array}
\newcolumntype{P}[1]{>{\centering\arraybackslash}p{#1}}
\usepackage{float}
\usepackage{booktabs}
\usepackage{xcolor}

\title{\centering \Huge AI Governance through Markets} 

\author{Philip Moreira Tomei$^{1,2}$, Rupal Jain$^{2,3}$, Matija Franklin$^{1,2,4}$}

\affiliation{
  \institution{AI Objectives Institute$^{1}$}
  \country{USA}
}
\affiliation{
  \institution{The ML Alignment \& Theory Scholars (MATS) Program$^{2}$}
  \country{USA}
}
\affiliation{
  \institution{Mercatus Center, George Mason University$^{3}$}
  \country{USA}
}
\affiliation{
  \institution{University College London$^{4}$}
  \country{UK}
}

\email{philip.tomei@aiobjectives.org}

\begin{document}

\begin{abstract}
  {\large This paper argues that market governance mechanisms should be considered a key approach in the governance of artificial intelligence (AI), alongside traditional regulatory frameworks. While current governance approaches predominantly focus on regulation, we contend that market-based mechanisms offer effective incentives for responsible AI development. We examine four emerging vectors of market governance: insurance, auditing, procurement, and due diligence, demonstrating how these mechanisms can affirm the relationship between AI risk and financial risk while addressing capital allocation inefficiencies. While we do not claim that market forces alone can adequately protect societal interests, we maintain that standardised AI disclosures and market mechanisms can create powerful incentives for safe and responsible AI development. This paper urges regulators, economists, and machine learning researchers to investigate and implement market-based approaches to AI governance.}
  
\end{abstract}

\maketitle
\thispagestyle{plain}  
\pagestyle{empty}  





{
  \setlength{\parindent}{15pt} 
  \setlength{\parskip}{0pt}    
  \tableofcontents
}
\newpage 

\section{Overview}

The field of Artificial Intelligence (AI) governance has predominantly emphasised regulatory frameworks \cite{anderljung2023frontier, schuett2023scope, roberts2021chinese} and international cooperation \cite{cha2024iaea, gruetzemacherinternational} to address AI risk. Meanwhile, uncertainty regarding AI risks is a major barrier to widespread enterprise adoption \cite{kpmg} \cite{altintas2024navigating}. Despite the economic benefits, organisations recognise AI risk as a business risk and lack the tools to confidently address it \cite{deloitte_2024}.  Market governance approaches, such as insurance, auditing, procurement, and due diligence, can serve to both mitigate AI risk and enable AI growth -  aligning market forces with prosocial behaviour \cite{smith2016art}. 

Market governance mechanisms are processes that structure economic behaviour by aligning financial incentives with desired outcomes. By directing capital flows, they possess the distinct advantage of embedding their own enforcement and incentive structures \cite{cullenward2020climate}. Regulatory initiatives for AI, on the other hand, have faced increasing criticism for being perceived as anti-competitive \cite{moutii2024europe} and anti-growth \cite{weinstein2024uk}. We contend that rational and responsible approaches to AI governance can align with economic objectives. Policy interventions may prove instrumental in creating a robust market governance ecosystem \cite{kiely1998neoliberalism}.

Analysis of the market governance of AI opens up distinct opportunities for both public and private involvement. Mechanisms such as insurance, auditing, due diligence, and procurement offer opportunities for both startups and large enterprises to capitalise on the growing need for AI de-risking which is projected to reach a value of \$276 billion by 2030 \cite{juniper2024assurance}. These mechanisms also afford policymakers and quasi-regulatory entities—including industry consortia, trade associations, and standards organisations — strategic pathways for market shaping. Through the deployment of incentive structures, subsidisation programs, public-private collaborations, and standardisation frameworks, policymakers can leverage markets to govern artificial intelligence development while advancing critical economic, technological, and societal imperatives.

\subsection{Governance Through Markets}

\begin{figure}[ht]
    \centering
    \makebox[\textwidth][c]{%
        \includegraphics[width=\textwidth, height=480pt]{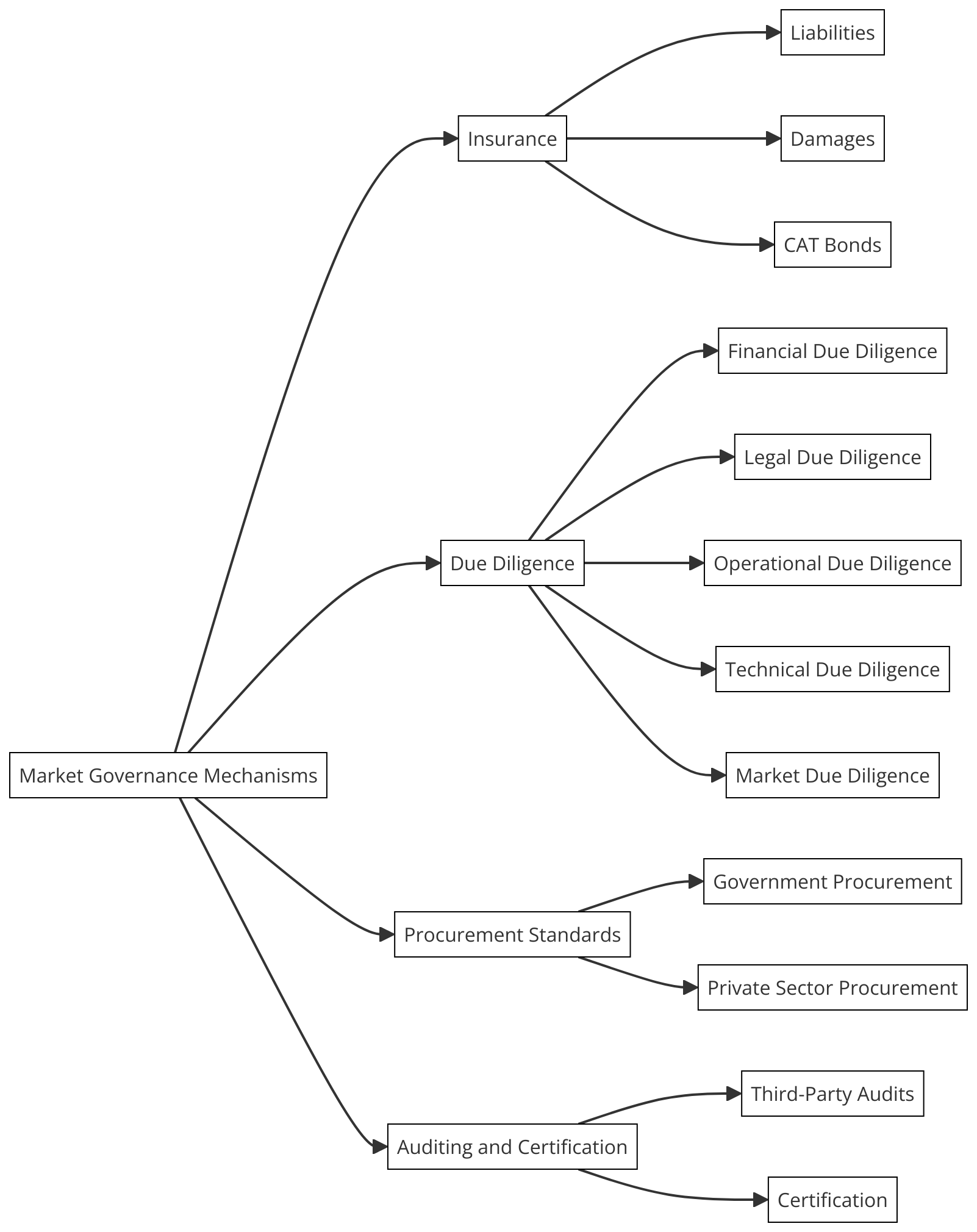} 
    }
    \caption{ 4 AI market governance mechanisms.}
    \label{fig:diagram1}
\end{figure}

By market governance mechanisms, we refer to tools and processes that align financial incentives with desired social or economic outcomes by structuring economic behaviour \cite{sbragia2000governance}. These mechanisms include (but are not limited to) insurance, auditing, procurement practices, and due diligence - which in turn serve risk distribution, information discovery, protocolisation and capital allocation. In the context of AI governance, market governance mechanisms aim to decrease AI risk by pricing, measuring and mitigating the risk incurred in AI deployment and development. These mechanisms further enable public and private stakeholders to shape AI in ways that align with broader societal and economic goals, fostering a balance between innovation and risk mitigation.

As Hadfield and Clark argue \cite{hadfield2023regulatory} AI fundamentally disrupts traditional regulatory frameworks in ways that demand comprehensive reconsideration of our approach.  Their analysis suggests that the primary challenge is not simply creating new rules for AI-specific risks, but developing dynamic regulatory approaches that can adapt to AI's pervasive influence. Their proposal for regulatory markets can be seen as one possible structure within which market governance mechanisms can operate. Their example demonstrates the broad array of policy strategies  available towards market governance of AI, which can range from complete self-governance to  government mandates with punitive enforcement.

Market governance mechanisms often function as private regulators, for-profit and non-
profit organizations that develop and supply regulatory services, which  compete to sell to targets, this is the case for instance with the protocol organisations and assurance providers \cite{micheler2020regulatory} we describe later.  This is not necessarily always the case, due diligence and insurance have similar effects but would not fall under this label. 

It is important to note that AI Risk and its associated uncertainty are a major barrier to AI adoption across the economy \cite{Ameye_Bughin_Zeebroeck_2023, mckinsey2024state}. This is exemplified by how AI service providers have resorted to providing unlimited indemnities to maintain their sales volume \cite{Service_Specific_Terms}. Market governance mechanisms work to reduce asymmetry and risk, enhancing the stability and efficiency of markets and enabling adoption \cite{kauffman2021mechanisms}.  For example, insurance functions as market governance by facilitating the distribution and management of uncertainty across a broad pool of market participants \cite{knight1921risk}, thereby mitigating individual exposure to risk while promoting resource allocation through the collective absorption of losses \cite{boettke2021economics}.

We do not believe the aforementioned four mechanisms to be exhaustive but rather exemplary of the kind of intervention points which are able to work alongside market forces to govern AI. We wish to highlight the way markets are already governing AI and widen the possibility space for AI governance research beyond hard regulation. Each of these mechanisms represents an opportunity for both public and private interventions into the governance of AI backed by the power of markets.

\subsection{AI Risk as Market Failure}

 The uncertainty around AI risk is a major barrier to enterprise adoption of AI technologies  \cite{kpmg, altintas2024navigating}. Large enterprise customers are uncertain about how to measure, mitigate and govern the risks posed by AI integration into their organisations \cite{deloitte_2024}. Notwithstanding its potential economic advantages, this issue persists as a significant impediment to the dissemination of AI technologies. Consequently, organisations have already recognised that AI-related risks are synonymous with business risks, and are responding in a manner that reflects this understanding. The awareness of these risks and the inability of organisations to address them prevents firms and society at large from benefiting from AI technologies. 

In this vein, we may understand the uncertainty around AI Risk and its mitigation as market failure. The failure, that is, to measure and mitigate the  private costs of AI development and deployment and price its externalities. By AI risk, we imply a risk from the development and deployment of AI systems. This risk to the firm is magnified in how closely AI is being integrated to value-generating and infrastructural operations. We use the following formula to quantify this risk-adjusted value \cite{artzner2007coherent} of AI investments:

\[
\text{RAV} = E(X) - \lambda \cdot \sigma(X)
\]

\begin{itemize}
    \item \textbf{RAV (Risk-Adjusted Value)}: The value of an outcome adjusted for risk, reflecting the market actor’s risk tolerance. It represents the certainty-equivalent value of the expected return after accounting for the investor's aversion to risk.

    \item \textbf{\(E(X)\) (Expected Value)}: The average projected financial return, calculated as the probability-weighted mean of all potential outcomes. Here, \(E(X)\) reflects the private expected returns for a market actor and does not inherently account for externalities unless they are explicitly included in the calculation. 

    \item \textbf{\(\lambda\) (Risk Aversion Coefficient)}: A factor representing a market actor’s sensitivity to risk, where higher values denote a stronger preference for certainty and a greater aversion to risk. It scales the impact of risk (as measured by \(\sigma(X)\)) on the risk-adjusted value, personalising the adjustment based on individual or organizational risk preferences. 

    \item \textbf{\(\sigma(X)\) (Standard Deviation)}: A measure of outcome variability, indicating the level of uncertainty or risk in financial performance. It quantifies the dispersion of possible returns around the expected value \(E(X)\), serving as a proxy for risk or volatility. A higher standard deviation signifies greater uncertainty and potential variability in returns.
\end{itemize}

This formula calculates a risk-adjusted value (RAV) by subtracting a risk penalty ($\lambda \cdot \sigma(X)$) from the expected value. While \(E(X)\) represents the average projected return, \(\sigma(X)\) quantifies the uncertainty or variability in potential outcomes. This separation allows for a more nuanced adjustment that reflects an actor's specific risk tolerance, represented by the coefficient \(\lambda\). Without this adjustment, the model would not adequately account for individual or institutional preferences regarding risk.

Imagine \(E(X)\) as the expected winnings from a lottery, averaged across all possible outcomes. While this gives a sense of the potential reward, it doesn't reflect the variability between a jackpot and losing the ticket price. \(\sigma(X)\) measures this variability, and \(\lambda\) adjusts for how much the individual dislikes the uncertainty of not knowing whether they'll win or lose. \(\lambda\) captures the market actor’s sensitivity to uncertainty, and \(\sigma(X)\) quantifies the variability in potential financial outcomes due to AI deployment. The contour plot of RAV in Figure 1 provides a visual representation of how \(E(X)\) and \(\sigma(X)\) influence the attractiveness of AI investment projects under varying \(\lambda\). In AI hype-cycles, \(\lambda\) is low, thus resulting in investment with higher \(\sigma(X)\) \cite{chong2004risk}.

Behavioural factors also play a crucial role in the accurate assessment of AI investment risks \cite{ackert2009behavioral}. During hype cycles surrounding AI technologies, cognitive biases may lead firms to underestimate their risk-aversion coefficient (\(\lambda\)) or the variability of outcomes (\(\sigma(X)\)). This is marked by the financial underperformance of many AI investments to date. Thus, investment decisions are not aligned with the market actor’s true risk tolerance \cite{rapp2014flexibility, baker2014investor}.

Furthermore, the RAV formula primarily focuses on financial variability and may overlook regulatory and ethical risks \cite{calo2017ai}. These include potential costs from regulatory fines, legal liabilities, or ethical breaches—such as violations of data privacy laws, biased decision-making algorithms, or unintended harmful consequences of AI deployment. Such risks are not always fully captured by \(\sigma(X)\) but can have substantial financial and reputational repercussions. Incorporating these factors into the risk assessment framework is necessary to obtain a comprehensive understanding of the risks involved in AI investments \cite{sharpe1964capm}.

Integrating the RAV formula into AI investment evaluations provides a structured approach to quantifying and managing the risks associated with AI deployment. This methodology reinforces the importance of acknowledging both private costs and externalities, leading to more responsible investment decisions. By properly accounting for the risk of AI development and employment, firms can better align their strategies with their actual risk tolerance levels. This alignment ensures benefits for shareholders through optimised financial performance and for stakeholders by mitigating negative societal impacts associated with AI technologies.

\begin{figure}[H]
    \centering
    \includegraphics[width=\textwidth]{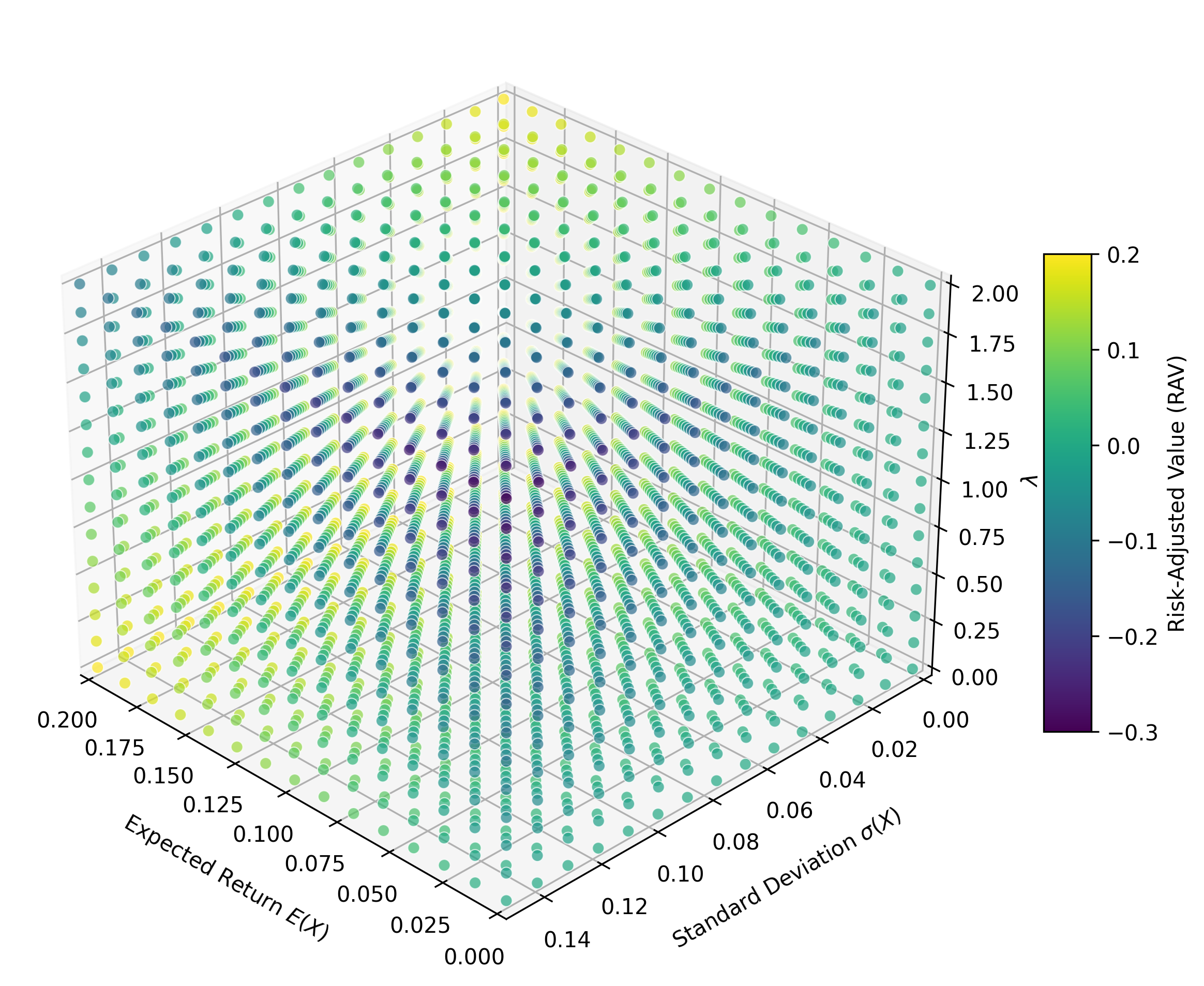} 
    \caption{Figure 1. D Scatter Plot of RAV as a Function of \(E(X)\), \(\sigma(X)\), and \(\lambda\). This visualisation illustrates how varying each parameter affects the RAV, with warmer colours indicating higher risk-adjusted values.}
    \label{fig:fig1} 
\end{figure}

As seen in Figure~\ref{fig:fig1}, at lower values of $\lambda$, the Risk-Adjusted Value (RAV) closely aligns with the expected return $E(X)$, as the risk penalty is relatively small. This reflects scenarios where risk aversion is minimal and the primary focus is on maximising returns, regardless of the associated volatility. Conversely, at higher values of $\lambda$, the risk term $\lambda \cdot \sigma(X)$ becomes more influential in the RAV calculation, causing RAV to decrease even when $E(X)$ is large. In risk-tolerant scenarios, where $\lambda$ is small, investments or decisions with higher volatility ($\sigma(X)$) may still yield favourable RAVs if $E(X)$ is sufficiently high to offset the modest risk penalty. This interplay highlights the importance of calibrating $\lambda$ to reflect the appropriate level of risk tolerance for a given context.

\section{Risk Distribution: Insurance}

Insurance serves as an enabler of transactions amid radical uncertainty by transferring and mitigating risks associated with unforeseen events \cite{knight1921risk}. In the context of deploying AI systems in economic functions, there is an unavoidable material financial risk due to the probabilistic nature of modern AI models \cite{ghahramani2015probabilistic}, as model outputs are undetermined by their inputs. As AI systems become more complex, agentic, and autonomous, the potential for unintended consequences  increases and so do prospective financial losses.

Insurance can serve as a powerful governance mechanism, shaping responsible AI development and deployment. This can be achieved through risk-based pricing, underwriting standards, and carefully crafted policy exclusions. It also offers potential tools to address thorny liability issues surrounding AI, such as joint causation scenarios or damages caused by hacked AI systems. While traditional insurance domains may exhibit relative stability, AI systems present uniquely complex risk profiles characterised by both substantial magnitude and high variance, which are characteristics that traditionally catalyse insurance market development \cite{lior2020ai}. The convergence of high-stakes outcomes and inherent unpredictability in AI agent behavior creates natural market conditions for novel insurance products, reflecting the insurance sector's historical role in facilitating technological adoption through systematic risk distribution.

Insurance has the potential to create a more favourable incentive structure by making practices such as “safety-washing” or underestimating AI-related risks less appealing \cite{tomei2024commodification}. The rigour of actuarial assessments may engender an anti-speculative framework for pricing AI development, curtailing so called 'AI Hype' - much like it has done in other sectors, such as real estate in climate-vulnerable areas \cite{nyce2015capitalization}. By pricing, mitigating, and transferring risk, insurance can encourage more accurate assessment and analysis of AI risk profiles. In so doing, the development of a robust AI insurance market may enable both first- and third-party stakeholders to more accurately understand the risk profile of AI development.

Cyberrisk provides a crucial precedent for AI risk. This has emerged as the most pressing form of business risk, mirroring the indispensability of software in every major enterprise \cite{wef2021risks}. Accordingly, he financial toll of cyberattacks swells by 15\% annually up to \$23tn by 2027 \cite{economist2024cyber}. Among numerous potential solutions to this pervasive insecurity, insurance offers some respite. A robust cyber incident insurance market incentivises organisations to strengthen their cyber defences, ultimately reducing societal cyber risk \cite{marsh2021cyberclaims}. While still in its infancy, cyberrisk insurance has promoted accurate analysis and pricing of the risk landscape. It further incentivises the adoption of cybersecurity measures by making plain the financial consequences incurred and reducing premiums or disqualifying policyholders with insufficient security measures.

As AI becomes integrated into the value creation of major industries, AI insurance will be in increasing demand. Lior et al\cite{lior2021insurance} have emphasised the value of building upon existing insurance infrastructure rather than creating entirely new AI-specific policies, given uncertainties in the field. AI will affect multiple lines of existing insurance, including Technology Errors and Omissions (E\&O)/Cyber Liability, Professional Liability, Media Liability, and Employment Practices Liability, among others, depending on the specific use case of the AI \cite{aon2024risks}. The manner in which AI risk is incorporated into insurance frameworks has significant implications for both insurers and policyholders, shaping the management of AI-related risks throughout the private sector.

Societal-scale AI incidents may prompt the development of a CAT-bond market, in order to transfer liability from the insurance sector to capital markets. As systemic AI risk increases with adoption this may provide a key backstop for societal resilience. Catastrophe bonds decrease insurers' contributions to systemic risk by offloading liability to global capital markets, thus lowering default risks and enhancing stability\cite{c42f96c98609cbef5c040a29f2a6b6263801a295}. CAT bonds support faster economic recovery after disasters by providing rapid liquidity and mitigating financial strain on issuers, as strongly supported by empirical studies \cite{74aedba9f56f14095656c362690da861ef83250f}. As the Crowdstrike incident shows, misalignment between parametric triggers and actual losses remains a significant cause of basis risk \cite{cc0ca731db738c8e06e3c8d71b54099cce528349}. 

Despite these potential benefits, and  a number of commercial AI Insurance ventures already underway \cite{munich2024insurance}, numerous issues still arise in AI risk distribution. The opacity of many AI systems ("black box" characteristics) presents severe technical challenges in vulnerability assessment. Furthermore, the widespread integration of  shared AI components, particularly foundational model APIs, creates potential cascading failures across industries, complicating the evaluation of systemic risks, a concern that is particularly acute in high-stakes sectors such as finance and healthcare.  Privacy risk assessment faces trade-offs between model accuracy and data protection, compounded by legal ambiguity around PII classification - while current fairness metrics often misalign with legal standards, increasing discrimination liability \cite{munich2024insurance}. Intellectual property risk is similarly constrained by varying technical and legal interpretations, while tort law uncertainty complicates risk assessment. As such, a number of regulatory, legal and scientific innovations are still necessary for a robust AI risk market to develop. 

\subsection{Case Study: Commercial Real Estate}

Commercial real estate (CRE) owners have faced increasing insurance premiums in recent years, prompting behaviour changes to manage these rising costs. Insurance companies have raised their prices to reflect the heightened risks associated with properties in high-risk areas, such as those prone to natural disasters \cite{greystone2024risk}. In response, property owners updated valuations to reflect accurate replacement costs, made significant property improvements (e.g., installing new roofs), and opted for higher or aggregate deductibles to negotiate better rates. These improvements reduced the overall risk of property damage, which in turn lowered the likelihood of claims. This demonstrates how a market mechanism—higher insurance premiums—successfully incentivised risk-reducing behaviours, ultimately benefiting both insurers and property owners through reduced risk \cite{bain2024future}. Additionally, alternative risk transfer strategies, such as structured programs, provide fixed premiums for budget certainty. These actions helped CRE owners reduce premiums and ensure greater financial stability \cite{greystone2024risk, bain2024future}. A market actor—a company—changed their behaviour to reduce risk and secure better insurance premiums.

\section{Assurance: Auditing and certification}

Third-party auditing is an independent examination of a company's financial statements, operations, or compliance by an external auditor to ensure accuracy and adherence to established standards and regulations \cite{kueppers2010audit}. Certification often follows auditing, serving as formal recognition that a company meets specific industry benchmarks or quality standards. These processes are usually made public, creating incentives for companies to maintain high levels of transparency and integrity to build trust with stakeholders, including investors, customers, and regulators. The public nature of auditing and certification encourages companies to improve their practices and reduce risks, as failing to meet standards can lead to reputational damage or financial penalties. Consequently, companies may change their behaviour by implementing stronger internal controls \cite{kueppers2010audit}

Audits, when viewed as market governance mechanisms, serve as tools that help regulate market behaviour by enhancing transparency, accountability, and trust between market participants \cite{reider2007financial}. They play a crucial role in reducing information asymmetry, as the US Chamber of Commerce \cite{doty2014audit} notes, reinforcing market confidence through reliable audits is critical for encouraging long-term investment.

There is significant literature on AI Auditing. The motivation for Audits for this varies; some \cite{ee2023cybersecurity, bernardi2024societal} argue that auditing is necessary for public safety, while others \cite{manheim2024necessity} further point towards keeping AI firms accountable to their own commitments. As Manheim et al. note,  findings from audits, when disclosed, provide regulators and the scientific community with critical insights into system behaviours and limitations. This transparency supports regulatory initiatives and scientific advancements by deepening the comprehension of system operations. Additionally, Auditing aids due diligence and procurement processes by providing third-party verification to first-party claims. 

Further, audit mechanisms facilitate the resolution of public disclosure  trade-offs, while system opacity maximises security at the expense of external oversight, comprehensive disclosure enhances scrutiny but risks exposing proprietary technology. Audits potentially offer an intermediary solution, enabling independent risk assessment while maintaining security protocols \cite{sharkey2024causal}. Successful audit completion enhances investor confidence and stakeholder trust, aligning institutional credibility with demonstrated accountability measures.

\cite{leech2024questionable} identify significant methodological concerns regarding research practices, particularly within private sector institutions, where incentive structures favor reporting superior performance on machine learning benchmarks. Their documentation of 43 distinct "questionable research practices"—actions that, while not constituting explicit research misconduct, materially misrepresent AI system capabilities—underscores the critical role of audit procedures in ensuring disclosure integrity and accuracy in AI system evaluation.

\subsection{Case Study: Zoom Video Communications}

Following its initial public offering, Zoom Video Communications, Inc. encountered intensified examination regarding security vulnerabilities within its telecommunications infrastructure, particularly during the unprecedented utilisation surge accompanying the COVID-19 pandemic \cite{bbc2020zoom}. Suboptimal encryption protocols and potential unauthorised data access vectors presented significant risks to stakeholder confidence and institutional reputation \cite{washington2020zoomsecurity}. 

In response, Zoom implemented comprehensive third-party audit and risk management protocols, engaging cybersecurity firms—notably Trail of Bits and NCC Group—to conduct thorough security assessments. These evaluations facilitated substantial enhancement of security infrastructure, including the implementation of universal end-to-end encryption protocols, materially improving service security parameters \cite{csa2022zoombreach}. This case study exemplifies the efficacy of third-party audits as market governance mechanisms for risk identification and mitigation.

\section{Protocolisaton: Procurement Standards}

Current procurement standards for AI systems are evolving to address challenges of safety, transparency, accountability, and security. Governments often rely on private contractors for AI development, necessitating careful planning during procurement to ensure transparency and compliance with norms \cite{Dor_Coglianese_2021}. Guidelines for AI procurement are being developed to support decision-makers in identifying key elements for secure AI systems, considering technical and regulatory environments \cite{Kieseberg_Buttinger_Kaltenbrunner_Temper_Tjoa_2022}. There is a particularly urgent demand for AI-specific public procurement   standards outside of the defence industry\cite{Hickok_2022}. Private organisations are further developing AI procurement standards, but concerns remain regarding liability and the need for internal governance frameworks \cite{Oketunji2023LargeLM, mckinsey2024state}. 

The U.S. military, as the world’s largest single customer, has an annual budget of \$916 billion \cite{statista_military_spending}, surpassing the total expenditures of most nation-states. This substantial spending is governed by stringent procurement standards, which, in turn, influence the behaviour of commercial suppliers seeking to secure highly lucrative government contracts  \cite{mccullough2024public}. Procurement standards, whether set by governmental bodies or large private entities and consortia, play a crucial role in shaping customer demand and, by extension, the development of products and technologies. 

These guidelines and principles function as bottom-up market governance mechanisms that effectively regulate industries while driving innovation \cite{decarolis2021buyers}. In the defence sector, procurement standards such as the Defense Federal Acquisition Regulation Supplement (DFARS) impose stringent requirements on suppliers, affecting not only contract eligibility but also driving technological innovation and quality improvements \cite{schwartz2008defense}. The Department of Defense, governed by the Federal Acquisition Regulation (FAR) and DFARS, enforces safety, performance, and reporting requirements \cite{executive_services_2014}. They dictate the evolving technological and operational standards for defence contractors, from the initial design phase through development, testing, and final delivery to meet the evolving needs of the military, thereby indirectly promoting advancements in technology while imposing strict reliability thresholds  \cite{gao2020defense}. We must note however that the  substantial resources required for regulatory compliance create barriers to entry, potentially diminishing market competition \cite{markowski2010defense}. Moreover, the transnational nature of these procurement standards manifests in their influence on international suppliers, who must conform to these regulatory requirements to participate in the U.S. defense market, effectively establishing global procurement paradigms \cite{schmidt2018globalization}.

In the commercial sector, multinational corporations such as Walmart and Apple implement procurement protocols that significantly influence supplier practices across multiple domains, including labour conditions, environmental sustainability metrics, and product quality specifications, thereby exercising considerable authority over global supply chain architectures \cite{gereffi2016economic}. This phenomenon exemplifies how procurement standards transcend their conventional role as mere operational guidelines to function as instrumental mechanisms in shaping industry practices, driving innovation trajectories, and establishing de-facto regulatory frameworks.

Pre-market testing standards, such as the FDA’s 510(k) Premarket Notification process, offer a compelling parallel for how large-scale buyers can use their purchasing power to shape the trajectory of AI development. Just as the 510(k) process expedites regulatory approval for medical devices that demonstrates substantial equivalence to another legally U.S. marketed device, procurement standards for AI could require companies to meet predefined risk assessment benchmarks in order to reduce the friction of entering into new lucrative markets and procurement contracts. This would incentivise a minimum level of efficacy and safety while also fostering innovation within a controlled framework. In the downstream, this may harmonise high AI standards across borders as well. In addition, pre-market testing standards typically come with public accountability measures, including the obligation to report adverse events or safety concerns post-market. Such transparency creates a feedback loop, encouraging developers to be proactive in mitigating risks while under the scrutiny of investors, regulators, and the public -- as discussed further in the section on “AI disclosures.”

The U.S. The Department of Defense's Cybersecurity Maturity Model Certification (CMMC) exemplifies existing technical safety standards for AI, delineating three-tiered cybersecurity requirements—basic protection, Controlled Unclassified Information (CUI) safeguards, and advanced threat mitigation—with mandatory compliance verification. This framework can inform the development of comprehensive AI procurement standards. However, the establishment of universally applicable AI safety and reliability metrics is imperative for practical implementation.

\subsection{Case Study: NASA Apollo Mission}

The National Aeronautics and Space Administration's (NASA) Apollo Programme (1960s-1970s) exemplifies how rigorous procurement protocols can facilitate the successful execution of complex, high-risk initiatives \cite{ertel1978apollo}. This program substantially enhanced both private sector competition and aerospace technological reliability while maintaining safety parameters in mission-critical contexts \cite{sperber1973apollo}.

NASA's procurement framework is structured upon the Federal Acquisition Regulation (FAR) and its organisational supplement, the NASA FAR Supplement  (NFS). These regulatory mechanisms are specifically designed to address the distinctive requirements of space exploration, encompassing stringent reliability metrics, safety protocols, and performance criteria \cite{mccaffrey2021nasa}. NASA's uncompromising adherence to safety protocols, precision requirements, and performance criteria necessitates enhanced capabilities among contractors \cite{arrilucea2018apollo}. 

These protocols catalysed technological innovations that would establish industry benchmarks for subsequent decades, encompassing advancements in propulsion systems, microelectronic components, computational software, materials engineering, navigational systems, and solar energy technology—innovations that were subsequently adapted for civilian applications  \cite{ntrs1971apollo}. Contemporary commercial aviation and aerospace enterprises, including SpaceX and the Boeing Company, maintain comparably stringent safety protocols per NFS procurement standards  \cite{faa2017human_spaceflight}.

\section{Investor behaviour: Due Diligence}

Investor due diligence refers to a comprehensive appraisal conducted by investors to assess the potential risks and rewards associated with an investment opportunity. This process involves a systematic evaluation of financial statements, management competency, market conditions, and regulatory compliance \cite{snow2023mergers}. For economists, due diligence plays a pivotal role in addressing information asymmetry—the imbalance of information between corporate insiders and external investors—which can lead to suboptimal investment decisions and market inefficiencies \cite{Novikov2018DUEDE}. For investors, the ‘the goal of due diligence is to make the buyer comfortable enough to go through with the deal’ \cite{snow2023mergers}.

Investor due diligence mitigates information asymmetry by enabling investors to obtain and analyse information that may not be readily apparent from public disclosures \cite{daley2024diligence}. This process encourages issuers to maintain accurate and comprehensive records, knowing that sophisticated investors will scrutinise their disclosures. Thus, due diligence acts as a self-regulating mechanism that complements formal regulatory frameworks \cite{daley2024diligence}.

Due diligence is not a monolithic process but rather a multifaceted approach which includes: 

\begin{itemize}
    \item \textbf{Financial Due Diligence:} This involves scrutinising financial statements, cash flow projections, debt obligations, and tax liabilities. In the context of AI, this extends to evaluating the financial implications of AI investments and their potential returns \cite{howson2017diligence}.
    \item \textbf{Legal Due Diligence:} Examining contracts, intellectual property rights, pending litigation, and regulatory compliance. With AI, this includes assessing algorithmic bias, regulatory compliance, and potential liability issues arising from AI decision-making \cite{howson2017diligence}.
    \item \textbf{Operational Due Diligence:} Evaluating business processes, supply chain efficiency, and operational risks. In AI-driven companies, this includes assessing the robustness of risk management, development cycles and cybersecurity practices \cite{harvey1995diligence}
    \item \textbf{Technical Due Diligence:} For tech companies, this involves evaluating the underlying technology, its uniqueness, and its potential for obsolescence. In the AI context, this becomes even more critical, requiring specialised expertise to assess AI models, data quality, and algorithmic performance \cite{harvey1995diligence}. 
    \item \textbf{Market Due Diligence:} Analysing market size, growth potential, competitive landscape, and customer base. With AI, this extends to evaluating the potential disruptive impact of AI technologies on existing markets, harnessing standardised disclosures to compare AI integrations across competing firms \cite{howson2017diligence}.
\end{itemize}

 The analysis that arises in due diligence helps in accurately pricing securities, reflecting all available information. Its governance ability lies in being the determinant process by which investors allocate capital. In so doing, it incentivises management practices and accountability within firms \cite{schilling2021accountability} and shapes competitive dynamics between firms. It is important to note that during the due diligence process, buyers are not solely focused on identifying potential risks. Equally important to investors is the evaluation of the enterprise's strengths and opportunities. Thus, due diligence is both a carrot and a stick, rewarding those organisations that present the least risk for investors while promising attractive returns \cite{hudson2007guidance}. 

Significant uncertainty exists regarding around the financial risk of firms with high levels of AI adoption \cite{Ameye_Bughin_Zeebroeck_2023}. Due diligence can help scrutinise AI practices, in doing so investors help establish market norms for transparency in AI development. Companies are further incentivised to maintain high standards in operational practices and regulatory compliance, knowing that investors pay close attention to these aspects during the due diligence process \cite{charles2023blockchain}.

\subsection{Case Study: BP Oil Spill}

The 2010 BP Deepwater Horizon incident, resulting in the discharge of approximately 4.9 million barrels of oil into the Gulf of Mexico, represents the most extensive marine oil spill in recorded history \cite{pallardy_britannica}. Beythe catastrophic environmental and economic implications, BP incurred \$65 billion in remediation costs and legal settlements (United States Environmental Protection Agency, 2024). The financial impact was equally significant: BP's market capitalisation decreased by \$105 billion, representing a 55\% decline in equity value \cite{hargreaves2010bp}.

While operational safety deficiencies and insufficient corporate oversight precipitated the disaster, subsequent investor due diligence proved instrumental in risk mitigation and catalysed substantial improvements in corporate governance and safety protocols \cite{heineman2011bp}. Investor pressure compelled comprehensive disclosure of financial liabilities and risk exposure, both immediate and long-term. Consequently, BP allocated \$5 billion annually toward transitioning from high-risk hydrocarbon operations to renewable energy technologies, demonstrating how due diligence can facilitate strategic reorientation toward sustainable development \cite{ziady2020bp}. This case study illustrates how investor-mandated audit reports and enhanced disclosure requirements, particularly regarding high-risk applications, can fundamentally reshape corporate risk management strategies.

\section{Standardised Information as a Foundation for Market-Based AI Governance}

Asymmetry exists between the knowledge held by AI providers and rs and the broader market—including investors, insurers, and procurers. Additionally, substantial uncertainty exists for all market participants regarding AI systems\cite{Ameye_Bughin_Zeebroeck_2023}. Without adequate information, markets are unable to make informed decisions \cite{healy2001information}. This information asymmetry and uncertainty, may be preventing firms from effectively allocating resources to AI \cite{Ameye_Bughin_Zeebroeck_2023}. 

All the market governance mechanisms we examine—including insurance providers, procurement entities, and due diligence evaluators—necessitate accurate and reliable data regarding artificial intelligence systems to effectively price AI assets, evaluate associated risks, and facilitate implementation protocols. These stakeholders require comprehensive information to ensure compliance with relevant standards, conduct thorough risk assessments, and optimise asset allocation decisions. We posit that the convergence of these requirements would be optimally addressed through the implementation of standardised disclosure. 

Standardisation of this kind in technology markets often reduces transaction costs, drives competition on value and reduces externalities \cite{Tassey}. Systematic disclosure could foster a robust epistemological foundation that enhances market participants' decision-making capabilities, ultimately enabling more efficient capital allocation and minimising societal risks associated with AI deployment \cite{mazzucato}.

\subsection{Disclosure Mitigates Information Asymmetry}

Information asymmetry significantly impacts investors' perceptions of novel technological implementation. It can result in both the underestimation and overestimation of \(\lambda\). In the former, market actors with less information on a technology might underestimate their aversion, acting as if they are more willing to accept financial risk. The risk aversion coefficient \(\lambda\) varies with the degree of information asymmetry \(I\):

\[
\lambda(I) = \lambda_0 \cdot e^{s_\lambda \cdot I}
\]

\begin{itemize}
    \item \(\lambda_0\): The \textbf{baseline risk aversion coefficient} when information asymmetry is zero (\(I = 0\)). This represents investors' standard level of risk aversion under conditions of full transparency.
    \item \(s_\lambda\): The \textbf{sensitivity parameter} indicating how the risk aversion coefficient \(\lambda\) responds to changes in information asymmetry \(I\). The sign and magnitude of \(s_\lambda\) determine the nature and extent of this response:
    \begin{itemize}
        \item \textbf{If \(s_\lambda > 0\)}: The risk aversion coefficient \(\lambda\) \textbf{increases} exponentially with increasing \(I\). This means that as information asymmetry grows, investors perceive higher uncertainty and become \textbf{more risk-averse}. They may \textbf{overestimate their actual risk aversion}, acting more cautiously than necessary due to the lack of information.
        \item \textbf{If \(s_\lambda < 0\)}: The risk aversion coefficient \(\lambda\) \textbf{decreases} exponentially with increasing \(I\). In this scenario, as information asymmetry increases, investors perceive themselves as \textbf{less risk-averse}, potentially \textbf{underestimating their true risk aversion}. This can lead them to take on excessive risk, possibly due to overconfidence or underappreciation of the unknown risks.
    \end{itemize}
\end{itemize}

By incorporating information asymmetry into the risk aversion coefficient, we can model how varying levels of transparency affect investment decisions. The sensitivity parameter \(s_\lambda\) quantifies the impact of information asymmetry on investors' risk preferences. A positive \(s_\lambda\) suggests that increased information asymmetry leads to higher perceived risk aversion, potentially resulting in underinvestment due to overestimated risks. Conversely, a negative \(s_\lambda\) indicates that higher information asymmetry causes investors to act less risk-averse than they actually are, possibly leading to overinvestment in risky AI projects.

\begin{figure}[H]
    \centering
    \includegraphics[width=\textwidth]{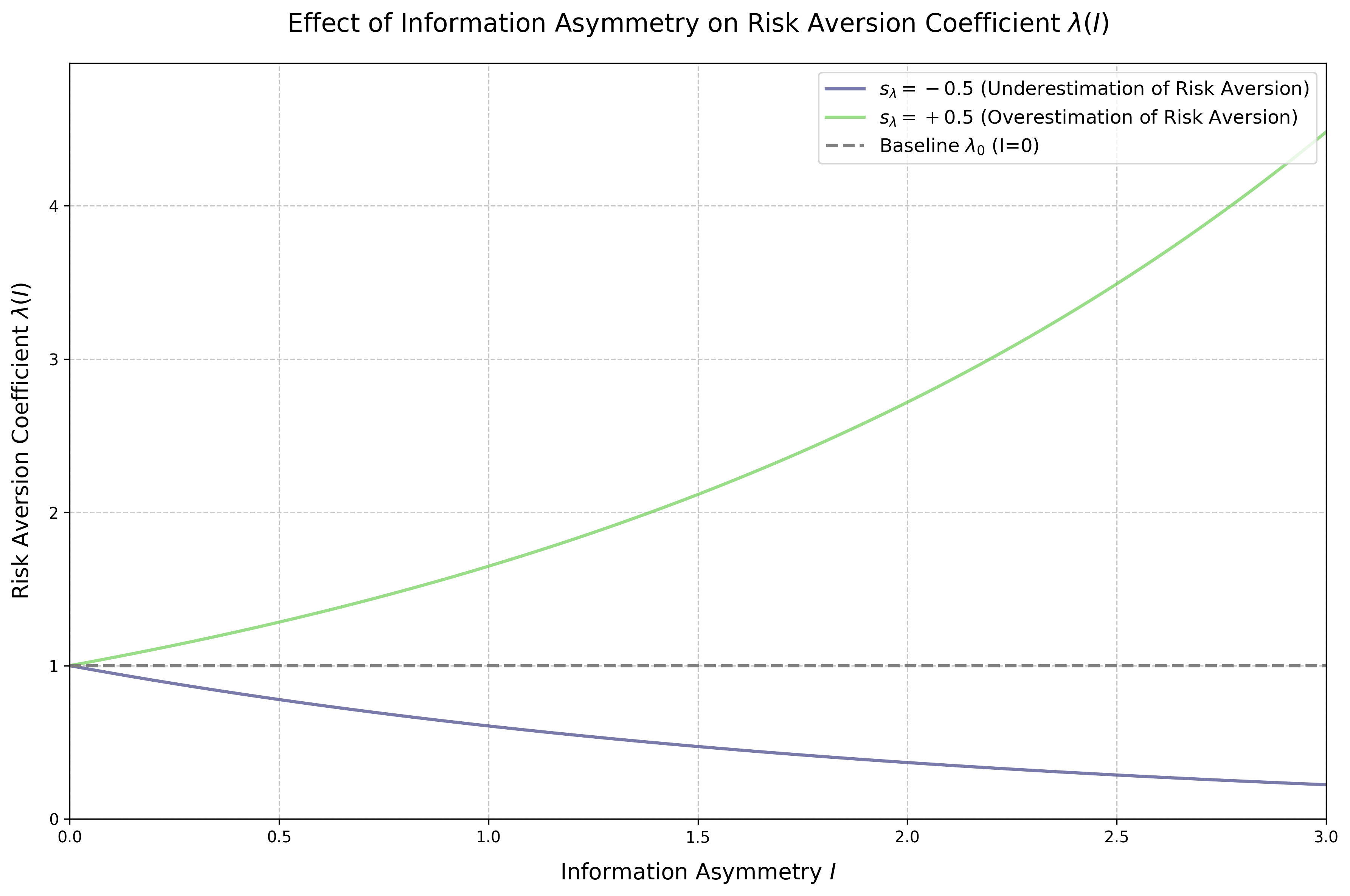} 
    \caption{Effect of Information Asymmetry on Risk Aversion Coefficient \(\lambda(I)\). The graph illustrates how \(\lambda(I)\) varies with information asymmetry \(I\) for \(s_\lambda = -0.5\) (Underestimation of Risk Aversion) and \(s_\lambda = +0.5\) (Overestimation of Risk Aversion).}
    \label{fig:fig2} 
\end{figure}

\section{Disclosing AI’s development and deployment}

Disclosure refers to the process of releasing all relevant information pertaining to a company. It is the foundation of the aforementioned market governance mechanisms as disclosure powers auditing, insurance, due-diligence and procurement. Disclosure can be financial or non-financial \cite{vanstraelen2003disclosure, musweu2018disclosure}. Central to financial disclosures are financial statements—comprising the balance sheet, income statement, and cash flow statement—which collectively provide insights into a firm’s financial position, operational performance, and liquidity \cite{hassan2019measurement}. Non-financial reporting encompasses the disclosure of a company's operations, business risks, and corporate strategies that are not captured in traditional financial statements \cite{turzo2022nonfinancial}. 

Most non-financial disclosures are not obligatory, yet 1,395 investors, who collectively manage \$59 trillion—accounting for 60\% of global assets under management—have endorsed the UN Principles of Responsible Investment \cite{pri2024annual}. This endorsement signifies a commitment to, among other considerations, take non-financial disclosures into account when making investment decisions. Some funds will include or exclude potential investments on the basis of their non-financial information thus creating a strong incentive \cite{curtis2021esg, dikolli2022esg}. There is also large investor demand for non-financial disclosure \cite{thewissen2024esg}. For example, it is common for investors to prefer funds that do not invest in companies providing services to the military \cite{starr2008sri}. Instances of widespread, non-financial disclosures encompass research and development (R\&D) information \cite{koh2014rd}, sustainability reports \cite{saini2022disclosures}, and organisational practices \cite{arvidsson2011disclosure}. Crucial for our purposes, nonfinancial disclosures provide market actors with information on potential risks that may not be immediately apparent in financial statements \cite{pozzoli2022nonfinancial, flostrand2006valuation}. 

\subsection{AI Disclosures}

Standardisation serves to simplify complex risk areas, rendering them accessible to a broad range of market participants who may lack expertise in AI.  By adopting a common framework, AI risks can be articulated in a manner that facilitates more informed decision-making across investment, insurance, and procurement activities. The absence of a standardised approach to reporting AI-related risks introduces significant challenges for investors, insurers, and other stakeholders. Without uniform metrics, assessing the financial and operational implications of AI risks becomes difficult, resulting in inefficiencies in capital allocation and decision-making processes. A standardised risk framework is essential, as it provides consistent and comprehensible metrics for communicating complex AI risks to non-specialists within the market. While such frameworks may not fully capture the intricacies of each risk, they serve as a practical means of ensuring that AI-related concerns can be communicated and addressed effectively.

We argue that the AI research communities can contribute to AI governance by standardising the disclosure of AI. This standardisation requires the consistent description and quantification of key aspects of AI systems to enable firms, policymakers, and other actors to make informed decisions. 

Providers, such as developers of foundational general-purpose AI systems like GPT-4o, create the underlying models, while distributors use these models to offer products, such as ChatGPT \cite{gutierrez2023gpa}. A growing community of researchers in academia, civil society, government, and industry is actively studying the risks associated with AI. These risks are not always negative, but rather descriptions of various features of the AI models and organisations that develop or distribute them that are currently uncertain. These risks are increasingly communicated through various market governance mechanisms. For instance, organisations such as METR, which function as safety auditors, evaluate the risk profiles of companies such as Anthropic or OpenAI \cite{achiam2023gpt4}. Over this decade, the standardisation of how AI-related information is transmitted through markets will likely solidify, much as we already see with performance benchmarks such as GPQA \cite{rein2023gpqa}, MATH \cite{hendrycks2021math}, and MMLU \cite{hendrycks2020massive} being widely adopted by companies like OpenAI, DeepMind, and Anthropic.

In a competitive environment, the pressure to reduce risk becomes even more pronounced. AI companies that successfully minimise risk can gain a competitive edge by attracting capital, securing better terms, and winning more procurement contracts. This competitive advantage can create a virtuous cycle where reducing risk not only enhances a company’s market position but also encourages industry-wide adoption of safer practices. As competitors observe the benefits of risk reduction, they too will be motivated to improve their risk profiles, leading to a broader industry shift towards safer AI development and deployment.

By establishing clear guidelines for describing risks— such as the allocation of compute, incident reporting, and capabilities in various areas—organisations can provide a comprehensive view of their AI models' vulnerabilities and limitations. This would be paired with specific metrics to quantify these risks, such as model error rates, data privacy breach incidents, or energy consumption during model training and deployment. Such quantification enables stakeholders to make informed comparisons between different AI systems and providers. With standardised reporting, market actors can assess AI risks more accurately, reducing uncertainty and promoting better capital allocation. 

The aforementioned market governance mechanisms require a framework for AI disclosure that identifies different sources of risk. To disclose requires an understanding of the different areas that are present in the development and deployment of AI, as well as in the organisations that develop and deploy. For example, OpenAI's preparedness framework identifies categories such as: 1) Individualised persuasion, 2) Cybersecurity, 3) Chemical, biological, radiological, and nuclear threats, and Autonomous replication and adaptation \cite{openai2023preparedness}. Many other such frameworks exist, with the AI Risk Repository capturing 700+ risks extracted from 43 existing frameworks \cite{slattery2024risks}. As such, this present paper will not be providing an exhaustive literature review of different AI Risks. 

For the purpose of standardizing AI Risk disclosure, two critical aspects must be emphasized. First, Risk should not be construed as inherently negative but rather as associated with uncertainty. A risk area is an area of uncertainty. Second, the way in which different aspects of AI are disclosed needs to be standardised in order for them to be utilised by and communicated between different market actors. This involves turning a complex nuanced area into a single metric or groups of metrics \cite{tate2007standardisation}. Some standardised metrics have already emerged, such as the size of models expressed with floating-point operations per second (FLOPS) \cite{chen2023flops}, and the generalised capability of AI being regularly disclosed with famous benchmarks such as GPQA, MMLU and MATH \cite{rein2023gpqa, hendrycks2020massive,  hendrycks2021math}. More metrics will be needed in order to fully describe the vast range of potential applications of AI.

\begin{figure}[H]
    \centering
    \makebox[\textwidth][c]{%
        \includegraphics[width=1.4\textwidth]{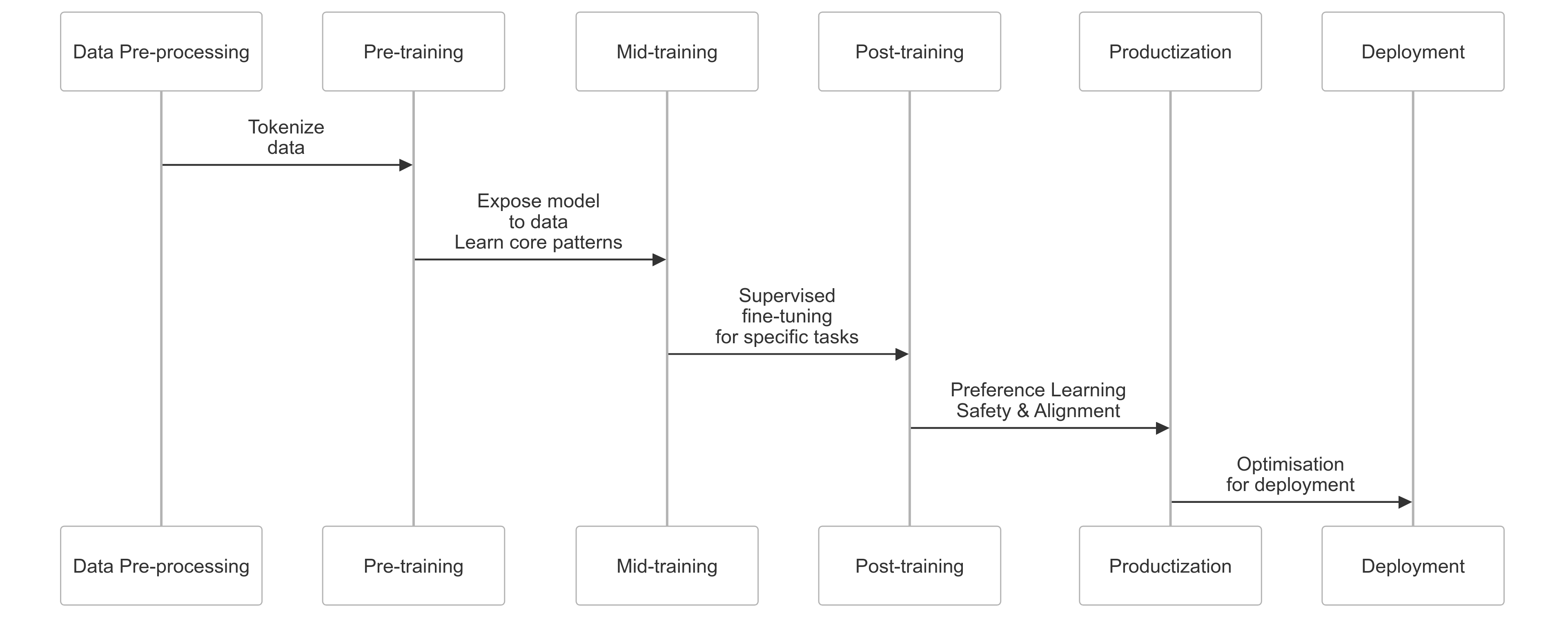} 
    }
    \caption{The 6 macro stages of the large AI model production line. This is a simplified overview. Other stages are described in the article text. This is not a comprehensive overview of large AI model development.}
    \label{fig:diagram2}
\end{figure}

Standardised metrics of AI risk will describe different aspects of AI development and deployment. The development of large models—encompassing large language models, reasoning models, multimodal models, and other advanced AI systems—follows a production pipeline with clearly delineated stages, each of which plays a critical role in shaping the model's capabilities (See Figure~\ref{fig:diagram2}) \cite{dubey2024llama}. The process initiates with data acquisition, where diverse datasets are curated and selected, which are subsequently subjected to tokenisation and preprocessing \cite{liu2024understanding}. Pre-training follows, where models are exposed to extensive datasets, learning foundational patterns \cite{du2024stacking}. It is succeeded by mid-training, a phase that includes fine-tuning for specific tasks, domain adaptation, and multi-task learning \cite{parthasarathy2024ultimate}. Post-training includes reinforcement learning from human feedback (RLHF) and other techniques, as well as additional safety work such as audits, red-teaming, and capability enhancement like teaching tool use \cite{ouyang2022training, wei2024jailbroken, mazeika2024harmbench}. Upon completing these stages, models are prepared for productisation, where they are optimised for deployment environments, such as through quantisation or distillation to reduce computational load  \cite{wang2024art, sreenivas2024llm}. Deployment involves integrating models into production systems, accompanied by continuous monitoring to track performance and detect issues like data drift \cite{jang2024driftwatch}. 

Within this production pipeline, a number of aspects are of material significance to market participants, including investors, regulators, and insurers. A detailed understanding of the risks, methodologies, and safeguards associated with each stage is essential for accurately assessing the potential returns and liabilities inherent in the development of large models. Enhanced transparency throughout these processes can facilitate improved risk assessment, inform capital allocation decisions, and ensure compliance with regulatory frameworks, thereby supporting models that meet stringent safety, ethical, and performance benchmarks.

Table~\ref{tab:table1} outlines some standardisation targets that are anticipated to become prevalent across various stages of the large model production process (See Appendix for a more detailed description of each standardisation target). Companies are already disclosing some of these metrics. These projections are grounded in current regulatory requirements, industry best practices, and emergent trends in AI governance. It is important to note that this is not an exhaustive inventory. Future research should extend this, providing a more comprehensive perspective on the requisite standards for developing responsible and market-aligned models.

\begin{table}[H]
\centering
\begin{tabular}{|p{4cm}|p{5cm}|p{5cm}|}
\hline
\textbf{Risk Area} & \textbf{Description} & \textbf{Standardisation Target} \\ \hline
\textbf{Data Acquisition and Privacy in AI Systems} & Diverse datasets are built by scraping, buying, or making data. This involves data curation (e.g., filtering out low-quality or biased data) \cite{liu2024understanding}. & Data provenance, detailing the sources and methods of data collection—whether through scraping, crowdsourcing, or third-party purchases—and providing transparency on labelling practices \cite{el2022impossible, carlini2022quantifying, ganji2019approaches}. \\ \hline
\textbf{Energy: Data Centre Efficiency and Algorithm Design} & Large AI models have large energy demands, driving the need for energy-efficient infrastructure and algorithms \cite{chong2013data, khan2022workload, mittal2014power, mo2021energy, zeng2023energy, patterson2021carbon}.  & Energy consumption metrics like Power Usage Effectiveness (PUE), the proportion of energy derived from renewable sources, the energy sources they use, training methods reducing computational burden \cite{jaureguialzo2011pue}.  \\ \hline
\textbf{Compute Allocation} & Allocation of compute resources to the training processes and the safety measures applied throughout AI model development \cite{dubey2024llama, liu2024understanding}. & Disclosing the quarterly compute spend on pre-training, mid-training, post-training, safety work, and inference. \\ \hline
\textbf{Intended Model Behaviour} & Large model developers shape their model's behaviour through techniques such as RLHF \cite{ouyang2022training}, Direct Preference Optimisation (DPO) \cite{rafailov2024direct}, fine-tuning \cite{parthasarathy2024ultimate}, red-teaming \cite{mazeika2024harmbench}, and output filters \cite{li2023unlocking}. & Report the intended model behaviours, and the methods and processes that are being employed to achieve these intended behaviours. The OpenAI Model Spec provides a strong example of transparency in this area \cite{openai2024model_spec}. \\ \hline
\textbf{Model Interpretability} & Model interpretability involves methods, such as \textit{Explainable AI} \cite{Linardatos2020ExplainableAA} and \textit{Mechanistic Interpretability} \cite{Conmy2023TowardsAC}, that enhance the transparency and understanding of AI systems \cite{Gunning2019DARPAsEA}.  & Interpretability techniques, model architecture, identified biases, Foundation Model Transparency Index \cite{Bommasani2023TheFM}. \\ \hline
\textbf{Open-source practices} & Open-source practices in AI development allow developers to share code, datasets, and methodologies \cite{Shrestha2023BuildingOA}. & Documentation of open-source contributions, including code, datasets, and methodologies. Disclosing incidences of potential dual-use cases, governance structures for contributions and policies for IP rights and licensing. \\ \hline
\textbf{Adversarial Testing} & Adversarial testing assesses the resilience of AI systems by exposing them to challenging inputs designed to exploit vulnerabilities \cite{ruiz2022adversarial, hannon2024resiliency}. & Detailing adversarial risks identified during testing, reporting the scale, frequency, and outcomes of these tests, as well as the data sources. \\ \hline
\end{tabular}
\caption{Risk Areas and Standardisation Targets}
\label{tab:table1}
\end{table}

\section{Discussion}

\subsection{Intellectual Property and Disclosure: A Tradeoff}

Artificial intelligence enterprises maintain substantial opacity regarding their operational processes, a phenomenon particularly pronounced in the semiconductor sector where competitive pressures and industrial espionage present significant concerns \cite{constantin2024integration}. However, empirical evidence suggests the possibility of achieving equilibrium between transparency and proprietary protection. Patent systems exemplify such balanced disclosure mechanisms \cite{brookings2011patent}. Standardised globally, patent legislation mandates comprehensive technical documentation of innovations, ensuring replicability by qualified practitioners, as codified in Article 83 of the European Patent Convention. This reciprocal arrangement confers temporary market exclusivity in exchange for the dissemination of public knowledge.

The pharmaceutical industry provides an instructive paradigm for managing the tension between disclosure requirements and intellectual property protection. Global pharmaceutical research and development expenditure exceeded \$300 billion in 2023, representing significant growth from \$5 billion in 1980 and \$38 billion in 2000 \cite{statista2024pharma}. Leading pharmaceutical corporations, including Merck, Novartis, and Roche, maintain substantial R\&D investment ratios, ranging from 27\% to 50\% of revenue \cite{pharmatech2024spending}. Their disclosure protocols prioritise methodological transparency over proprietary formulations \cite{deloitte2024accounting}, encompassing clinical trial methodologies, regulatory milestones, development timelines, R\&D intensity metrics, and scientific uncertainties, while maintaining competitive advantages through strategic patent protection \cite{harvard2021biopharma}. This disclosure framework enables investors to assess future profitability while establishing realistic expectations. The model has facilitated substantial capital attraction despite inherent developmental risks, supported by the potential for significant long-term revenue from successful therapeutic interventions.

The pharmaceutical sector's approach offers valuable insights for AI disclosure frameworks \cite{accesstomedicine2020trust}. Strategic transparency—encompassing risk categorization, regulatory compliance milestones, and mitigation protocols—can be achieved without compromising proprietary algorithms. This methodology, focusing on process disclosure rather than technical specifications, effectively balances regulatory compliance, investment attraction, and stakeholder trust. Thus, AI disclosure requirements may be reconceptualised as strategic opportunities rather than regulatory burdens.

\subsection{Market incentives for risk mitigation can unlock funding for safety research}

The scale of global asset management presents significant potential for influencing artificial intelligence development trajectories. Current market data indicates substantial capital pools: hedge funds manage approximately \$5 trillion, private equity firms control \$5 trillion, sovereign wealth funds oversee \$11 trillion globally, while U.S. pension funds alone manage approximately \$12 trillion in assets \cite{sarwar2023hedgefunds, whiteside2024pensions, wilson2024sovereignwealth}. If even 1\% of assets under management incentivise responsible AI initiatives, this could could mobilise approximately \$330 billion—a sum that exceeds current philanthropic contributions to AI safety by several orders of magnitude, as exemplified by Open Philanthropy's \$46 million allocation in 2023 \cite{mcaleese2023aisafety}. Although not comparable, redeployment of this kind could fundamentally transform the AI safety and security landscape, enabling expanded research initiatives, enhanced compliance frameworks, and accelerated development in interpretability and model safety protocols.

This potential capital reallocation can be contextualised against ESG mandates, where projections indicate asset managers will oversee \$33.9 trillion in ESG-focused funds by 2026 \cite{esr2024trends}. This precedent demonstrates institutional investors' capacity to integrate non-traditional criteria into investment decisions, despite uncertainty regarding ESG's correlation with financial returns. Crucially, the primary critique of ESG funds lies in their inconsistent metrics and lack of standardisation, which undermines comparability and transparency across firms and industries. This results in fostering superficial compliance practices, such as greenwashing, rather than driving substantive improvements in sustainability or ethical governance. Our argument for a standardised disclosure framework seeks to avoid an equivalent "safety washing". Further, the more direct financial implications of AI-related risks may present even more compelling incentives for institutional investors. Thus, strategic influence on capital flows could create substantial financial incentives for responsible AI development on a global scale.

\subsection{Scala Politica: Tailoring Intervention to Size without stifling innovation}

Disclosing information about AI Risk is a type of intervention, and to be effective every intervention must be tailored to the size of the entity that it is being applied to. Interventions applied to a country, may not apply to a city, and vice versa. Unwanted outcomes occur when an intervention designed for an entity of a certain scale, is applied to an entity of a different scale \cite{taleb2020principia}. The same can be said for the difference between the disclosure standards applied to startups and large organisations. 

The interconnected nature of market actors can create cascading effects throughout the market ecosystem. Institutional investors, particularly private equity firms managing pension fund portfolios, demonstrate how modifications in compliance expectations at upper market echelons can fundamentally reshape capital allocation patterns throughout the investment hierarchy. The dramatic 473.5\% increase in Fortune 500 companies identifying AI-related risks underscores this growing concern \cite{fortune2024ai_risks}. Conservative projections suggest that AI-related incidents resulting in merely a 2\% devaluation—whether through market capitalisation decline, regulatory penalties, or customer attrition—could precipitate substantial financial implications, given the magnitude of capital exposed.

However, startups have a starkly different landscape. The financial scale disparity between Fortune 500 corporations and venture capital markets is substantial, with the former generating collective revenues exceeding \$18 trillion in 2023, while venture capital investments totalled approximately \$394 billion in 2022, declining from a peak of \$611 billion in 2021 \cite{cfi2023fortune500}. Owing to the complexities inherent in AI supply chains, there are financial interdependencies between emergent enterprises and established corporations \cite{cahn2024aisupplychain}. Major corporations, notably cloud service providers, often mitigate the risks that startups face through substantial purchase commitments and preferential pricing strategies. While this arrangement currently maintains supply chain stability, its sustainability remains contingent upon the continued risk-absorption practices of technology conglomerates.

Startups face severe financial constraints where compliance costs can drastically affect their operating margins, as evidenced by research indicating that a 200\% increase in compliance costs could shift a startup's operating margin from 13\% to -7\% \cite{shorenstein2024compliance}. Growth of small and emerging enterprises is important for a dynamic economy, vibrant markets, and the preservation of a competitive spirit. Small businesses are underappreciated drivers of economic growth, contributing approximately 44\% of U.S. economic activity alone and creating 62\% of new jobs between 1995 and 2020 \cite{podium2024smb}. 

The imposition of extensive disclosure requirements on early-stage enterprises may impede innovation trajectories. Resource constraints can limit these organisations’ capacity to navigate these frameworks, resulting in increased operational costs that may deter market entry and extend commercialisation timelines \cite{nber2021regulationimpact}. Excessive regulatory oversight can foster risk aversion, potentially inhibiting experimental approaches essential for technological advancement \cite{dw2024overregulation}. Thus the appropriate level of disclosure must be adjusted to an entity's size. 

\section{Conclusion}

Our analysis demonstrates how market governance can function as instruments for promoting beneficial artificial intelligence development trajectories. These mechanisms represent critical intervention points through which both public and private entities can cultivate incentive structures that favour positive AI outcomes. Tools such as insurance, auditing, procurement, and due diligence, which have long played a role in managing risks and fostering transparency in sectors like finance and pharmaceuticals, are now beginning to emerge in the AI industry. Standardised disclosures are playing a growing role in reducing information asymmetry, enabling more informed decisions by investors, insurers, and regulators.

Our examination of market governance of AI reveals promising avenues for exploring AI risk reduction as a key to AI adoption. We encourage further investigation into these approaches by policymakers, economists, and machine learning researchers, as the intersection of their expertise may yield valuable insights for future AI governance interventions. Further research in this vein could significantly enhance our understanding of how market governance mechanisms might be optimally designed and implemented to ensure beneficial AI outcomes while maintaining innovation and economic efficiency.

\begin{acks}

We extend our deepest gratitude to Kristian Ronn, Ilan Strauss, Jamie Bernardi, Colleen McKenzie, Gabor Szorad and Alex Krasdomski for their contributions and invaluable insights to this research. 

We also thank Miles Brundage, Miguel Diaz, Nicolas Moes, Ema Provic, James Fox, Cyrus Hodes, Milan Griffes, Deric Cheng, Jacob Goodwin, Gabriel Weil, Lucy Philippon, Risto Uuk, Ryan Kidd, Shakeel Hashim, Deger Turan, Edmund Zagorin, David Lawrence and Oliver Klingfjord for their helpful support and feedback throughout this work.

\end{acks}

\bibliographystyle{ACM-Reference-Format}
\bibliography{Markets}

\appendix

\section{AI Risk Standardisation Targets}

\subsection{Data Acquisition and Privacy in AI Systems}

Data acquisition is a foundational step in developing large models, as the quality and diversity of datasets directly influence the performance and reliability of AI systems. This stage requires careful attention to privacy regulation especially given the legal landscapes (e.g., GDPR). As such, standardised reporting in this domain might include metrics related to data provenance, detailing the sources and methods of data collection—whether through scraping, crowd sourcing, or third-party purchases — and providing transparency on labelling practices. Recent legal challenges, such as lawsuits against AI companies for using copyrighted materials \cite{reuters2024anthropic, copyright2023courts} underscore the importance of curating datasets to exclude unauthorised content. A core issue in data privacy relates to how personal information is managed and protected during and after collection. High-performance models often require substantial data retention, which can inadvertently include sensitive information \cite{el2022impossible}. Private data, such as contact numbers and email addresses, can remain within training datasets and be extracted from models \cite{carlini2022quantifying}.

Companies can provide standardised disclosures on their data practices. This could include descriptions of data minimisation efforts, encryption standards, data anonymisation techniques, data storage practices, retention periods, and the purposes for which data is used. Disclosures may also cover details on third-party access to data, as well as how companies handle data subject requests, such as access, rectification, or erasure of personal information. Transparent reporting on data breaches—such as the number of sensitive records exposed, their nature, and mitigation measures—could further aid market participants in assessing a company’s risk management capabilities. Where appropriate, companies can employ existing standardisation metrics from frameworks such as the ISO/IEC 27001 for information security \cite{ganji2019approaches}, and the NIST Privacy Framework \cite{hiller2017privacy}.

\subsection{Energy: Data Centre Efficiency and Algorithm Design}

Large AI models have large energy demands, driving the need for energy-efficient infrastructure and algorithms. Data centre efficiency methods aim to optimise the energy use of computational resources to support model training. This includes strategies for aligning workload demands, resource allocation, and energy balancing \cite{chong2013data}, alongside software-oriented approaches for workload management \cite{khan2022workload}, and integrated power management strategies \cite{mittal2014power}. Simultaneously, algorithm design plays a critical role in improving the energy efficiency of machine learning processes, with techniques like federated edge learning \cite{mo2021energy}, hardware acceleration \cite{zeng2023energy}, and sparsely activated deep neural networks \cite{patterson2021carbon} offering significant reductions in energy consumption.

Standardised reporting on data centre efficiency might include energy consumption metrics like Power Usage Effectiveness (PUE) and the proportion of energy derived from renewable sources \cite{jaureguialzo2011pue}. Companies could also disclose which energy sources they use. For algorithm design, companies could report on energy efficiency metrics related to training processes, detailing how different techniques reduce the computational burden. Frameworks like the GRI \cite{petera2015global}, ISO 50001 \cite{marimon2017reasons}, and The Green Grid \cite{jaureguialzo2011pue} can already provide some reporting guidelines. 

\subsection{Compute Allocation}

Describing the allocation of compute resources would reveal information about the production processes and the safety measures applied throughout AI model development. Market actors could understand a company's strategic focus. For example, companies could disclose the quarterly compute spend on pre-training, mid-training, post-training, safety work, and inference \cite{dubey2024llama, liu2024understanding}. This macro overview could market participants, such as investors performing due diligence, evaluate where a company places its resources and how it balances the goals of innovation, safety, and efficiency.

\subsection{Intended Model Behaviour}

Large model developers use a series of different method to shape their model's behaviour. This involves shaping model outputs to prioritise valuable behaviours while preventing unsafe, illegal, or unethical content. Techniques such as RLHF \cite{ouyang2022training}, Direct Preference Optimisation (DPO) \cite{rafailov2024direct}, fine-tuning \cite{parthasarathy2024ultimate}, red-teaming \cite{mazeika2024harmbench}, and output filters \cite{li2023unlocking} are commonly employed to align model outputs with desired outcomes to ensure models remain safe for public use. A great focus is placed on avoiding unwanted behaviours, which result in reputational risks and legal liabilities. Key areas of concern include avoiding, harmful content, such as suicide ideation \cite{bhaumik2023mindwatch}, "hallucinations" \cite{Kaddour2023ChallengesAA}, and bias in model outputs \cite{Oketunji2023LargeLM}.

To standardise disclosures related to intended model behaviour, companies could report on their intended model behaviours, and the methods and processes that they are employing to achieve these intended behaviours. The OpenAI Model Spec provides a strong example of transparency in this area \cite{openai2024model_spec}. It details how it wants its models to behave, whilst trying to align with specific ethical objectives, safety measures, and user needs, balancing user directives with broader considerations like legality and societal impact. 

\subsection{Model Interpretability}

Model interpretability involves methods that enhance the transparency and understanding of AI systems \cite{Gunning2019DARPAsEA}. Often, AI models function as "black boxes," making it difficult to understand the reasoning behind their predictions. \textit{Explainable AI} \cite{Linardatos2020ExplainableAA} and \textit{Mechanistic Interpretability} \cite{Conmy2023TowardsAC} offer techniques for clarifying the logic behind model outputs, helping stakeholders gain better insights into the model's decision-making process.

Companies could standardise the disclosure of interpretability techniques to provide insights into how specific decisions are made. They might also disclose aspects of model architecture, any identified biases or limitations. There are technical breakthroughs that need to be made for this to be fully possible, but setting standards also sets a research aim. Frameworks like the Foundation Model Transparency Index  \cite{Bommasani2023TheFM} could guide some of these disclosures.

\subsection{Open-source practices}

Open-source practices in AI development allow developers to share code, datasets, and methodologies \cite{Shrestha2023BuildingOA}. This openness can accelerate progress and enable efficient peer review to identify and rectify errors. Open-sourcing advanced AI models can also come with risks, such as model poisoning and dual-use concerns (e.g., repurposing models for harmful  applications) \cite{Eiras2024RisksAO}. Additionally, transparency can challenge intellectual property (IP) management and compliance with data privacy laws.

To standardise open-source practices, companies might disclose detailed documentation of their open-source contributions, including code, datasets, and methodologies. Disclosing incidences of potential dual-use cases can aid regulatory bodies in understanding associated risks. Companies could also outline governance structures for contributions and establish clear policies for IP rights (IPR) and licensing, particularly when integrating open-source elements into commercial products. Companies can use aspects of existing frameworks like the Open Source Initiative \cite{initiative2004opensource} for licensing, or the Linux Foundation’s OpenChain Specification \cite{Coughlan2020StandardizingOS} for open-source compliance.

\subsection{Adversarial Testing}

Adversarial testing assesses the resilience of AI systems by exposing them to challenging inputs designed to exploit vulnerabilities. These methods are essential for evaluating how AI models withstand malicious attempts, especially in security-sensitive applications \cite{ruiz2022adversarial}. They help identify risks such as data poisoning, evasion attacks, and extraction threats, which can compromise model integrity \cite{hannon2024resiliency}. 

Companies could standardise disclosures by detailing adversarial risks identified during testing. Reporting the scale, frequency, and outcomes of these tests, as well as the data sources, can provide stakeholders with insights into a model's robustness \cite{Croce2020RobustBenchAS}.

\section{Existing Disclosure Frameworks}

Lack of corporate transparency coupled with inadequate risk assessment by investors creates a precarious foundation for potential catastrophic outcomes. The 2008 collapse of Lehman Brothers serves as a reminder about the far-reaching consequences of inadequate information transparency and (public, investors’, and regulatory) oversight in the markets. Leading up to its bankruptcy, Lehman had amassed significant exposure to subprime mortgages, which were increasingly seen as high-risk investments.  As homebuyers defaulted on loans they couldn't afford, it sent shockwaves to Wall Street. This led to a global financial crisis and millions of job losses and home foreclosures. Despite warning signs, such as declining housing prices and rising mortgage delinquencies, many investors continued to back Lehman’s stock. This disaster could have been uncovered at a nascent stage with rigorous oversight and due diligence. Lehman also employed deceitful accounting practices known as "Repo 105," temporarily removing debt from its balance sheet - a clear information asymmetry and lack of transparency between Lehman's management and stakeholders.\\

Industries with high transformative potential and safety-critical environments—like defence, finance, biotechnology, semiconductor, petrochemicals, pharmaceuticals—are primed to adopt a culture of AI safety due to their existing security mindsets. In these sectors, AI integration has the potential to scale operations at an unprecedented rate, but this rapid expansion often occurs without responsible oversight, magnifying risks and accelerating crises. The table below demonstrates how risks and opportunities emerging from AI can be disclosed within the limits of already existing reporting frameworks/guidelines.

\begin{table}[H]
\renewcommand{\arraystretch}{1.3} 
\setlength{\tabcolsep}{4pt} 
\centering
\begin{tabular}{@{}p{2.8cm}p{3.8cm}p{7.8cm}@{}}
\toprule
\textbf{Existing Frameworks} & \textbf{Existing Categories} & \textbf{Integration of AI Risk and Opportunities into Disclosures} \\
\midrule
ISO/TC 276, Biotechnology &
Analytical Methods, Bioprocessing, Data Processing and Integration, Biobank and Bioresources, Predictive Computational Models & 
\textbf{Predictive Computational Models:} Model overfitting, poor generalisation, lack of ethical oversight, biased predictions. \newline
\textbf{Biobank and Bioresources:} Privacy and security concerns, genetic data leaks leading to discrimination, mismanagement in biosample cataloguing/storage, and degradation from system failures. \newline
\textbf{Analytical Methods:} Issues in data integrity, validation errors, algorithmic misclassification. \newline
\textbf{Bioprocessing:} Risks from flawed algorithms, quality control failures, bioproduction yield optimization destabilizing supply chains, and privacy compromises. \newline
\textbf{Data Processing and Integration:} Interoperability errors, annotation biases impacting research. \\ 
\midrule
Pharmaceutical Supply Chain Initiative (PSCI) &
Ethics, Human Rights, Health and Safety, Sustainability, Materials, Supply Chain Resilience & 
\textbf{Ethics:} AI-driven medical decisions, bias in diagnostics, price gouging, and automated fraud detection. \newline
\textbf{Human Rights:} Surveillance in labor practices, job displacement, discrimination. \newline
\textbf{Health and Safety:} Automation of hazardous materials handling, mental health implications, failure to predict emergencies. \newline
\textbf{Sustainability:} Efficient resource consumption offset by increased emissions and waste. \newline
\textbf{Materials:} Cyber risks in traceability of sourcing materials. \\ 
\midrule
European Federation of Pharmaceutical Industries &
Patient Safety and Clinical Trials, Data Privacy, Supply Chain Transparency, IP Protection, RnD Transparency, Corporate Responsibility & 
\textbf{Patient Safety and Clinical Trials:} Bias in algorithms skewing trial outcomes, misdiagnosis, privacy breaches, lack of accountability. \newline
\textbf{RnD Transparency:} Reporting AI in drug discovery, ensuring diverse patient group applicability. \newline
\textbf{IP Protection:} Risks in disclosing sensitive model details. \newline
\textbf{Supply Chain Transparency:} Over-reliance on AI logistics predictions creating vulnerabilities. \newline
\textbf{Data Privacy:} GDPR compliance and patient data security challenges. \\ 
\bottomrule
\end{tabular}
\caption{Biopharmaceuticals: Integration of AI Risks and Opportunities into Disclosures}
\label{table:biopharmaceuticals}
\end{table}

\begin{table}[H]
\renewcommand{\arraystretch}{1.3} 
\setlength{\tabcolsep}{4pt} 
\centering
\begin{tabular}{@{}p{2.8cm}p{4cm}p{7.8cm}@{}}
\toprule
\textbf{Existing Frameworks} & \textbf{Existing Categories} & \textbf{Integration of AI Risk and Opportunities into Disclosures} \\
\midrule
Securities and Exchange Commission (SEC) & 
Risk Factors, Management's Discussion and Analysis (MDandA), Financial Statements, Internal Control over Financial Reporting, Executive Compensation, Legal Proceedings & 
\textbf{Risk Factors:} Market volatility from algorithmic trading, biased lending models, model failure, privacy violations, data theft, market swings, dependency on market makers, lack of transparency and interpretability in model outputs, information asymmetry, losses due to inaccurate model predictions, algorithmic collusion, market manipulation, unpredictability, data quality risks. \newline
\textbf{MDandA:} AI impact on financial performance, operational efficiencies, and systemic vulnerabilities due to AI integration. \newline
\textbf{Financial Statements:} AI-related assets, liabilities from AI investments, impairment risks, potential liabilities from cyber breaches, platform engagement statistics post-AI integration. \newline
\textbf{Executive Compensation:} AI-driven metrics for performance evaluation and biases. \newline
\textbf{Legal Proceedings:} Regulatory scrutiny of AI practices, litigation risks related to AI misuse, discrimination claims, failure to comply with privacy laws (GDPR, etc.). \\ 
\midrule
Solvency II Directive (EU) & 
Quantitative Requirements, Governance and Risk Management, Solvency and Financial Condition Report, Supervisory Review Process & 
\textbf{Quantitative Requirements:} Capital requirements for model risk, investment into AI safety, model performance statistics. \newline
\textbf{Risk Management:} Cybersecurity threats, data dependency (inaccurate financial projections), AI oversight mechanisms, transparency and accountability issues for AI-driven decisions, biased lending models, risk prediction models, stress testing with AI models. \newline
\textbf{Solvency:} Market instability, algorithmic trading impacts, spread of misinformation, manipulation of user behavior, insufficient capital reserves due to unexpected AI-related losses. \newline
\textbf{Supervisory Review Process:} Regulatory scrutiny over AI practices, evaluation of AI risk management practices. \\ 
\midrule
Financial Stability Board (FSB) reports & 
Systemic Risk, Regulatory Framework, Macroprudential Policy, Financial Stability Assessments, Data Quality and Governance, Institutional Resilience, Market Conduct and Consumer Protection, International Cooperation & 
\textbf{Systemic Risk:} AI-induced market volatility, algorithmic trading risks, flash crashes, stress testing scenarios. \newline
\textbf{Macroprudential Policy:} AI impact on financial cycles, AI's role in economic forecasting, policy responses. \newline
\textbf{Financial Stability Assessments:} Risks in high-frequency trading or algorithmic trading (e.g., flash crashes). \newline
\textbf{Data Governance:} Data quality and privacy risks, algorithmic decision transparency, stock manipulation. \newline
\textbf{Institutional Resilience:} Cyber threats and operational vulnerabilities, financial shock absorption capability. \newline
\textbf{Consumer Protection:} Misleading information due to "AI-washing," biased lending models, data privacy concerns. \newline
\textbf{International Cooperation:} Global coordination on AI governance, cross-border data sharing. \\ 
\bottomrule
\end{tabular}
\caption{Finance: Integration of AI Risks and Opportunities into Disclosures}
\label{table:finance}
\end{table}

\begin{table}[H]
\renewcommand{\arraystretch}{1.3} 
\setlength{\tabcolsep}{4pt} 
\centering
\begin{tabular}{@{}p{2.8cm}p{4cm}p{7.8cm}@{}}
\toprule
\textbf{Existing Frameworks} & \textbf{Existing Categories} & \textbf{Integration of AI Risk and Opportunities into Disclosures} \\
\midrule
Initiative for Responsible Mining Assurance (IRMA) Standard & 
Environmental Stewardship, Labor Rights, Mine Closure and Rehabilitation, Planning for Positive Legacies, Economic Responsibility, Security Arrangements, Cultural Heritage (Indigenous Peoples' Rights), Supply Chain Management & 
\textbf{Supply Chain Resilience:} Cyber vulnerabilities, supplier diversification, diminished traceability due to opaque AI. \newline
\textbf{Cultural Heritage:} Over-exploitation and erosion of biodiversity and tribal culture, inadequate representation of Indigenous knowledge. \newline
\textbf{Environmental Stewardship:} Algorithmic bias underestimating biodiversity loss, improper waste disposal, and resource depletion due to poor automated predictions. \newline
\textbf{Labour Rights:} Safety risks from AI-driven equipment, tribal unemployment, inadequate resource allocation for emergencies, job displacement due to mining automation. \newline
\textbf{Mine Closure and Rehabilitation:} Automated detection of environmental impacts post-closure. \\ 
\midrule
Extractive Industries Transparency Initiative (EITI) & 
Revenue Reporting, Payments to Governments (taxes, royalties, fees), Licensing and Contracts, Production Data, Ownership Structures, Social and Environmental Impact, Revenue Management and Expenditure, Sub-national Payments & 
\textbf{Revenue Reporting:} AI inaccuracies in revenue forecasting, impacts on fiscal transparency, and mismanagement of public funds. \newline
\textbf{Payments to Governments:} Automation errors in tax compliance. \newline
\textbf{Licensing and Contracts:} Compliance risks, AI-driven data theft, and legal reasoning issues. \newline
\textbf{Production Data:} Provenance issues, poor resource allocation due to inaccurate predictions. \newline
\textbf{Social and Environmental Impact:} Job displacement risks, pollution (air, water, waste dumping), biodiversity loss, habitat destruction for migratory animals. \\ 
\midrule
International Petroleum Industry Environmental Conservation Association (IPIECA) & 
Governance, Health, Safety, Environment, Social Responsibility & 
\textbf{Governance:} Lack of accountability in AI decision-making, algorithmic decisions in supplier selection. \newline
\textbf{Climate/Energy/Environment:} Energy consumption from AI-driven operations, inefficiencies in climate risk modelling. \newline
\textbf{Biodiversity:} Accelerated destruction from AI tools. \newline
\textbf{Safety, Health, and Security:} Predictive systems for accident prevention, risks from autonomous systems (e.g., drones, robotics). \newline
\textbf{Social Responsibility:} Job displacement due to automation, biased decision-making algorithms. \\ 
\bottomrule
\end{tabular}
\caption{Extractive Industries: Integration of AI Risks and Opportunities into Disclosures}
\label{table:extractive-industries}
\end{table}

\begin{table}[H]
\renewcommand{\arraystretch}{1.2} 
\setlength{\tabcolsep}{4pt} 
\centering
\begin{tabular}{@{}p{2.5cm}p{3.5cm}p{7.5cm}@{}}
\toprule
\textbf{Existing Frameworks} & \textbf{Existing Categories} & \textbf{Integration of AI Risk and Opportunities into Disclosures} \\
\midrule
Carbon Disclosure Project (CDP) &
Governance, Strategy, Risks and Opportunities, Targets, Emissions, Energy, Environmental Performance (Climate Change, Water Security, Forests, Biodiversity), Supply Chain, Engagement & 
\textbf{Risks and Opportunities:} Resource optimisation, predictive maintenance, smart grid management, safety monitoring, operational risks from AI decision-making errors, cyber risk, grid dysfunction. \newline
\textbf{Targets:} Using AI to achieve SDG, green energy, waste management (e-waste generation from hardware upgrades), scaling AI responsibly. \newline
\textbf{Emissions:} Real-time data analytics, optimizing data centres' location based on energy sources. \newline
\textbf{Climate Change:} AI’s role in climate adaptation/mitigation, forecasting, carbon footprint of LLMs. \newline
\textbf{Water:} AI’s use in water resource management, sustainable practices. \newline
\textbf{Forests and Biodiversity:} Ecological modelling, satellite imagery analysis, biodiversity loss mitigation. \\ 
\midrule
NIST Risk Management Framework & 
Prepare, Categorize, Select, Implement, Assess, Authorise, Monitor & 
\textbf{Prepare:} Identify AI risks (bias, privacy, security), set acceptable residual risk levels, establish ethical guidelines. \newline
\textbf{Categorise:} Classify risks by severity and sensitivity. \newline
\textbf{Select Security:} Apply AI-specific controls for privacy, robustness. \newline
\textbf{Implement:} Integrate controls, train workforce, mitigate bias. \newline
\textbf{Assess:} Test AI robustness, detect bias, validate models. \newline
\textbf{Authorise:} Approve for use, verify compliance, document risks. \newline
\textbf{Monitor:} Continuously track AI risks, feedback into strategies. \\ 
\midrule
International Integrated Reporting Council (IIRC) & 
Organisational Overview, Governance, Business Model, Risks and Opportunities, Strategy and Resource Allocation, Performance, Future Outlook & 
\textbf{Financial Capital:} Metrics for costs, savings, risks, obsolescence. \newline
\textbf{Intellectual Capital:} Proprietary technologies, over-reliance risks. \newline
\textbf{Human Capital:} Safeguard liberties, ensure equality and safety. \newline
\textbf{Natural Capital:} Efficiency, carbon footprint of data centres. \newline
\textbf{Social Capital:} Mitigate misinformation, adhere to GDPR. \newline
\textbf{Future Outlook:} AI lab partnerships, R\&D revenue percentage, dual-use cases. \\ 
\bottomrule
\end{tabular}
\caption{Environment and Sustainability: Integration of AI Risks and Opportunities into Disclosures}
\label{table:environment-sustainability}
\end{table}

\begin{table}[H]
\renewcommand{\arraystretch}{1.2} 
\setlength{\tabcolsep}{4pt} 
\centering
\begin{tabular}{@{}p{2.5cm}p{3.5cm}p{7.5cm}@{}}
\toprule
\textbf{Existing Frameworks} & \textbf{Existing Categories} & \textbf{Integration of AI Risk and Opportunities into Disclosures} \\
\midrule
Nuclear Regulatory Commission (NRC) & 
License Application and Review Process, Operations, Safety Culture Policy Statement, Cybersecurity Regulations, Emergency Preparedness, Environmental Review, Quality Assurance Requirements & 
\textbf{Cybersecurity Regulations:} Cybersecurity threats exploiting nuclear controls, extortion risks, espionage, adversarial attacks on safety-critical systems, incident response plans. \newline
\textbf{Operations:} Monitoring algorithm performance, addressing system failures, unauthorised access risks, decision-making transparency issues. \newline
\textbf{Safety Culture Policy Statement:} Use AI to enhance safety systems, balance human oversight, train staff on AI safety. \newline
\textbf{Emergency Preparedness:} Reliance on AI for emergency response, risk of misinformation, quality of training data affecting predictions. \newline
\textbf{Quality Assurance Requirements:} Model validation, testing protocols, documentation. \\ 
\midrule
International Atomic Energy Agency (IAEA) & 
Safety of Nuclear Installations, Nuclear Security, Safety Assessment for Facilities and Activities, Management System, Emergency Preparedness, Radiological Safety, Radioactive Waste Management, RnD & 
\textbf{Safety of Nuclear Installations:} Address AI-related control failures, accountability in opaque AI systems. \newline
\textbf{Nuclear Security:} AI vulnerabilities in surveillance systems, bias in threat detection algorithms. \newline
\textbf{Management System:} AI’s role in reactor monitoring, training employees for AI integration, simulating emergency responses. \newline
\textbf{Radiological Safety:} AI errors in detecting radiation levels, efficient waste transport and logistics systems. \\ 
\bottomrule
\end{tabular}
\caption{Integration of AI Risks and Opportunities into Nuclear Disclosures}
\label{table:nuclear}
\end{table}

\end{document}